\documentclass[12pt]{iopart}
\usepackage[utf8]{inputenc}
\usepackage[section]{placeins}
\usepackage{rotating}
\usepackage{float}
\usepackage{graphicx}
\usepackage{amssymb}
\begin{document}

\title[Seismic noise at BNO] {Seismic Noise Background in the Baksan Neutrino Observatory}

\author{L~Naticchioni$^{1}$, N~Iudochkin$^{5}$, V~Yushkin$^{3}$, E~Majorana$^{1,2}$,  M~Perciballi$^{1}$, F~Ricci$^{1,2}$, V~Rudenko$^{3,4}$}

\date{\today}

\address{$^1$INFN - Sezione di Roma, I-00185 Roma, Italia}
\address{$^2$Universit\`a di Roma {\it Sapienza},  I-00185 Roma, Italia}
\address{$^3$Sternberg Astronomical Institute of Lomonosov Moscow State University (MSU), Universitetskii pr. 13, Moscow 119234, Russia}
\address{$^4$Institute of Nuclear Researches RAS, str. 60-Letiya Oktyabrya 7a, Moscow 117312, Russia }
\address{$^5$Institute of Geosphere Dynamics, Russian Academy of Science, RAS, Leninskii prosp. 38/6, Moscow 117334, Russia}
\ead{luca.naticchioni@roma1.infn.it}

\begin{abstract}
In this paper we report a study of the seismic noise measured in the underground Baksan Neutrino Observatory in the framework of the site qualification for third generation Gravitational Wave detectors like the Einstein Telescope \cite{ETdesign2011} and Cosmic Explorer \cite{CE}. The main spectral feature below 1 $Hz$ is the oceanic microseism, while for greater frequencies the measured horizontal and vertical accelerations approaches the Peterson low noise model. Using two synchronized seismometers we also studied the coherence in the microseismic band ($0.1-0.5Hz$) between three underground stations located at $0.3km$, $1.4km$ and $3.7km$ from the tunnel entrance. Finally, on the base of our measurements we evaluate the Newtonian noise contribution from the seismic noise background of body waves.
\end{abstract}

\pacs{91.30.Ye, 04.80.Nn, 91.30.Fn}

\section{Introduction}
\label{sec:intro}
Gravitational waves predicted in the context of the Einstein  theory of Relativity causes space to be stretched in one direction while simultaneously being squeezed in the perpendicular direction. The space-time strain, or change in length divided by length, induced by these waves can be up to ten thousand times smaller than the diameter of a proton. 
The relative change in length is so low, $\Delta L/L \lesssim 10^{-21}$, (corresponding to a displacement sensitivity of  $\sim 10^{-18}$ m over a distance of $1$ km) that a gravitational wave (GW) detector must be extremely sensitive to the strain change and shielded from several sources of noise as \textit{e.g.}\ the ground motion, wind effect, lightning strikes, and in general seismic noise  that can mimic or hide a gravitational wave signal. \\
The environmental noise is attenuated by enclosing the sensitive detector in ultra high vacuum chambers and adopting active and passive techniques.  Both types of isolation increase the frequency bandwidth of the detector but, because of   the huge gap between the residual r.m.s. motion of the ground ($\sim 10^{-7}$ m)  and the displacement sensitivity of the detector, non-linear effects can limit significantly the performances and  the design of the active control loops is a challenge. In addition, the seismic noise in the microseismic band (from 0.1 to hundreds of mHz) can couple to the residual tilt of suspended elements in the GW detector, disturbing the control loops of these elements. Moreover, the seismic random motion of the ground around the gravitational test masses results in a stochastic gravitational force acting directly on them, and subsequently produces an additional  noise at the detector output. The latter is the so-called Newtonian Noise (NN) and poses a limit to the detector sensitivity in the frequency range below $\sim 5$  Hz. It is a direct gravitational coupling between the interferometer test masses and the surrounding medium in which compressional seismic waves propagates, generating mass density fluctuations. It is evident that this source of noise cannot be shielded or mechanically filtered. Therefore, finding a suitable quiet seismic environment is the first step in reducing these effects. \\

\noindent As part of the design study of a new generation of GW detectors, as the Einstein Telescope\cite{ETdesign2011}, we studied the seismic noise characteristics of various sites \cite{Amman}. In particular, we explored  underground locations that yield significant reduction in seismic power spectral density (PSD) in comparison with the sites hosting on surface the current GW interferometric detectors, like LIGO and Virgo.
These PSD measurements are usually compared with the Peterson new high (NHNM) and low (NLNM) noise models \cite{Peterson}, to give an empirical reference to the minimum or maximum seismic noise background that we can expect in a ground-based site on Earth.

\noindent In this paper we report the results of our measurement campaigns performed in the Baksan Neutrino laboratory. Two series of data have been collected with different instruments at distance of few years. In the following sections,  after a presentation of the Baksan laboratory and the geological structure of this geographic area, we summarize the results obtained in   2013 and in  2018. Then we conclude with a Newtonian noise estimation given the measured seismic background at the site.

\section{The Baksan Neutrino Observatory and its rock composition}

The Baksan Neutrino Observatory (BNO) is a scientific laboratory of the Institute for Nuclear Research of the Russian Academy of Sciences (INR RAS) located in the the Caucasus mountains of the Karbardino-Balkaria Republic in Russia. It started operations in 1977 \cite{kuzminov}, hosting mainly  neutrino experiments already at the USSR epoch.
The laboratory itself was built in a 4 km  long horizontal tunnel under the Mount Andyrchi, 4,000 m high. The entrance is at 1,700 m  from sea level, in the gorges of the Baksan river.
BNO was conceived to carry on studies and experiments of both fundamental and applied physics. The Observatory includes  surface installations for cosmic ray physics and underground laboratories for neutrino physics physics and physics of rare processes: it host also the Optoacoustic GRavitational ANtenna, OGRAN \cite{OGRAN}.\\

\noindent 
Geologically, the Caucasus Mountains belong to the Alpide belt system that extends from southeastern Europe into Asia; it is a border between the two continents and one of the world's higher mountain chain. The Greater Caucasus Mountains are mainly composed of Cretaceous and Jurassic rocks with the Paleozoic and Precambrian rocks in the higher regions.  The laboratory entrance is located in the gorge of the Baksan river,  not far from the Mont Elbrus (5642 m), a dormant volcano. This is the largest Quaternary volcano in the European part of the Russia, situated within the central part of Greater Caucasus. North of the Greater Caucasus the deep sedimentary Terek and Kuban foreland basin (more than $6000$ m thick; up to 1,600 m elevation) forms the transition to the Scythian platform. North-West of Mount Elbrus, the Stavropol ``high" forms a basement uplift. In the North Caucasus and in particular in basin of Baksan river, the volcanic and plutonic rocks are interpreted as parts of a single magmatic system with an age of 2.8-3.0 Ma, similar to the magmatic system of the molybdenum-bearing {\it Questa} caldera complex of New Mexico.  \cite{Bogatikov}

\begin{figure}[h!]
\centering
\includegraphics[width=0.8\textwidth]{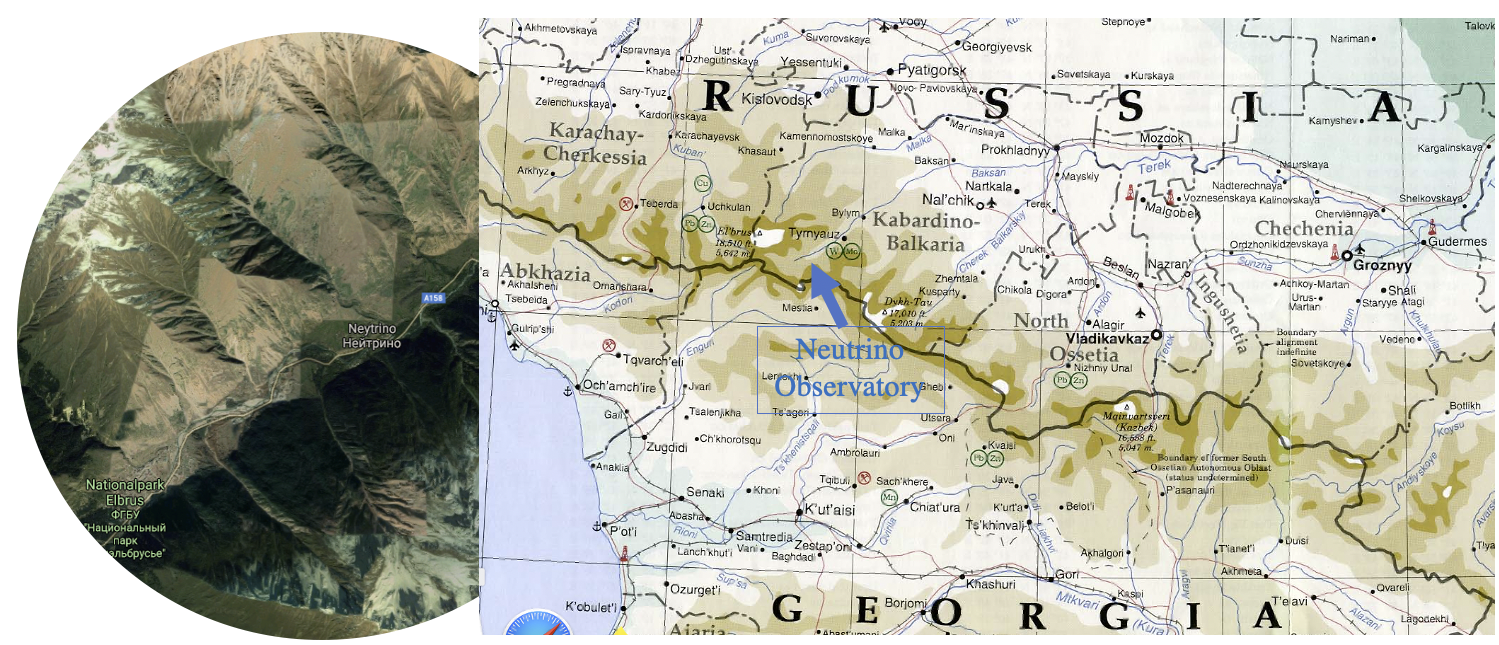}
\caption{The location of the Baksan Neutrino Observatory - BNO}
\label{fig:figure1}
\end{figure}

\noindent The surrounding rock geological minerals of the laboratory location are the plagiogneiss down to the depth 800 m, and plagio granites at more deep levels.  The main rock of the massif is shale, which has a thorium and uranium radioactive concentration close to that of granite. A density of these rock species practically is equal in average to $\rho=2800$ kg/m$^3$. The chemical composition of the local rocks is reported in \cite{Pomanski}.


 \begin{figure}[h!]
\centering
\includegraphics[width=1\textwidth]{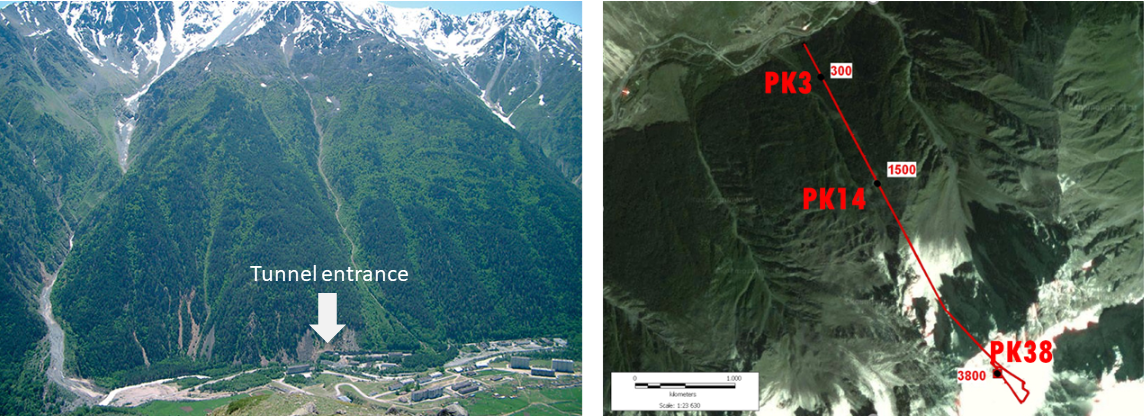}
\caption{Left: front view of the Andyrchi mount with the BNO tunnel entrance. Right: tunnel path on the satellite view of the BNO area, with the studied locations indicated}
\label{fig:BNOview}
\end{figure}

\begin{figure}[h!]
\centering
\includegraphics[width=0.8\textwidth]{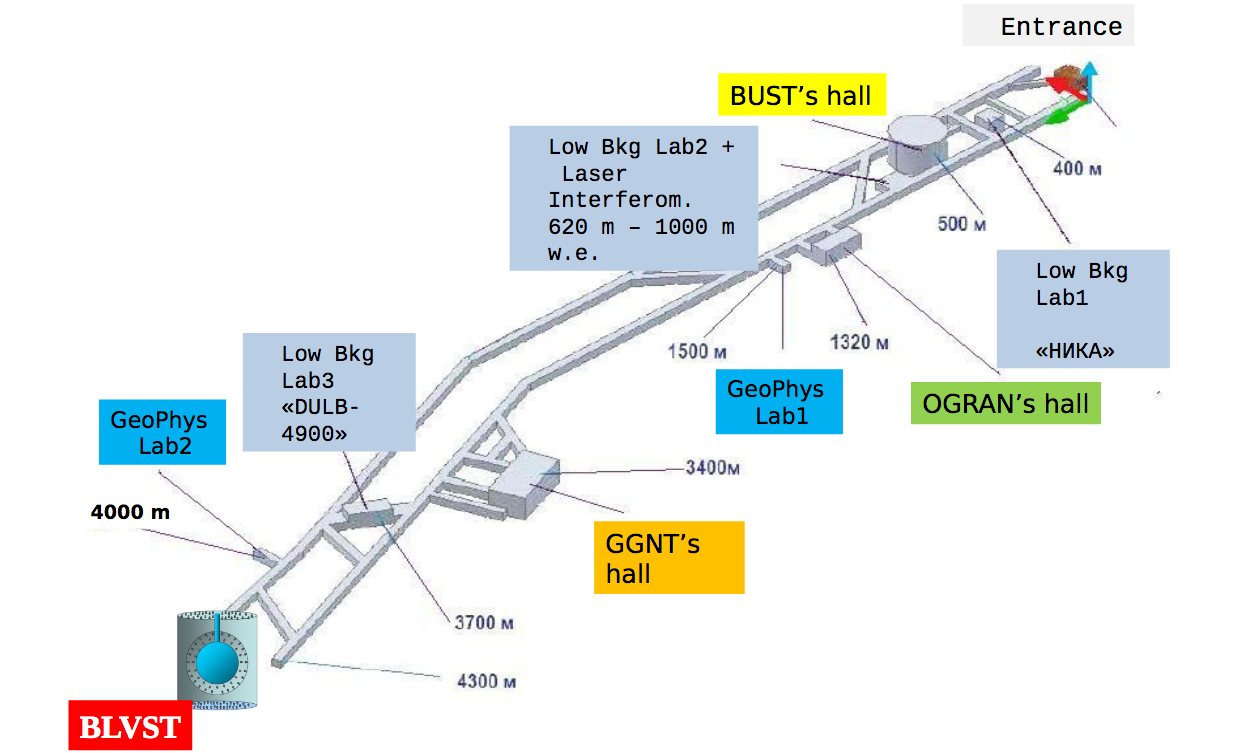}
\caption{Laboratories hosted along the BNO tunnel. The depth from the surface increases with distance along the tunnel from the entrance (right). Image credit: Y. Gavrilyuk}
\label{fig:BNOscheme}
\end{figure}

\noindent The underground laboratories, conceived to search for rare processes predicted  in theories of  elementary particle physics and  cosmology, are   distributed along the tunnel and they differ in their volume and depth. However, all  laboratories have been designed to have the special feature of a reduced  background caused by a surrounding radioactivity.
\noindent The experimental areas can be divided into two groups: the first one is the group of moderately deep zone with the largest one reserved for scintillation telescope laboratory. The second one is the group of very deep experimental areas including the gallium-germanium solar neutrino laboratory. The OGRAN experiment is hosted in a dedicated room located almost at half the away of the horizontal tunnel\cite{OGRAN2}.

\section{The 2013 data taking}\label{sec:2013}
In 2013, the measurements were carried out by the SAI MSU group with the assistance of GSRAS (Geophysical Service of the Russian Academy of Sciences \cite{RussianAcademy}), which has a stationary underground laboratory at $\sim 3800$ m from the tunnel entrance of the BNO main tunnel, equipped with a full range of geophysical meters. Investigations of the seismic background along the main tunnel were carried out with a portable seismometer within the EU program F7 (ASPERA grant entitled "Einstein's Telescope" \cite{ETdesign2011}).  
In 2013 measurements have been taken  with a conventional seismometer Guralp CMG-3T/ESPC, an instrument consisting of three sensors, which can measure the north-south, east-west and vertical components of ground motion simultaneously. The seismometer is equipped with a built-in digitiser, and it operates in the frequency range $(3 \times 10^{-3} ~-~ 50)$ Hz.
The measurement campaign have been carried on by monitoring the seismic background noise in three different locations along the 4 km-long horizontal tunnel, named  PK3, PK14, PK38.

\begin{figure}[h!]
\centering
\includegraphics[width=1.0\textwidth]{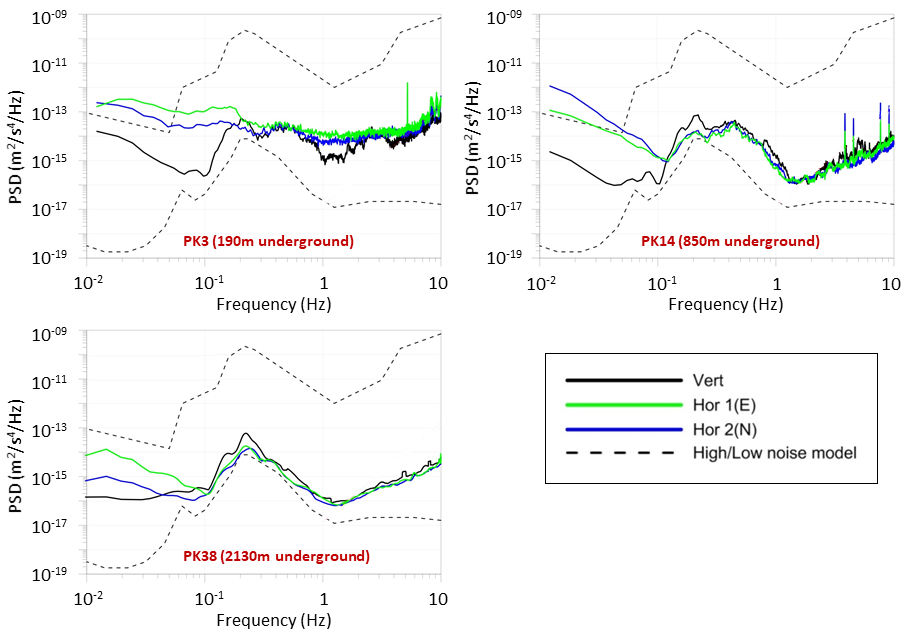}
\caption{Horizontal (green and blue lines) and vertical (black line) acceleration averaged power spectra at PK3 (top left), PK14 (top right), PK38 (bottom left) measured at BNO in 2013, compared to NLNM and NHNM (dotted lines)}
\label{fig:2013spectra}
\end{figure}

\noindent PK3 is the station  $\sim 300$ m far from  the tunnel entrance and with a depth of $ \sim 190$ m from the daily surface.

\noindent PK14 is  at a  distance of $\sim 1500$ m from the entrance of the main tunnel: the instrumentation was  in  the vicinity  of  OGRAN, the detector hosted in a facility  constructed using principles of solid-state and laser interferometer gravitational antennae.  The tunnel depth from the surface in this point is $\sim 850$ m. 

\noindent PK38 is  the station  at $\sim 3800$ m from the tunnel entrance, at a depth of  $ \sim 2130$ m; in this area  in the past there was  a geophysical complex with  tilt indicators, magnetometers, gravimeters, thermometers as well as earthquake detection stations devoted  to carry on a  geophysical survey.

Results of measurements are presented in figure \ref{fig:2013spectra} as a spectral density of horizontal and vertical acceleration in the logarithmic scale. At the deepest mark $3800$ m in the frequency range of $10^{-1}$ Hz, the acceleration spectral density decreases from $\sim 1\cdot 10^{-14}~~ m^2/s^4$ Hz  to $\sim 1\cdot 10^{-16} ~~m^2/s^4$ Hz.

\section{The 2018 data taking}
\label{sec:2018}
The second data taking  was carried on  during three days in 2018, from September $25^{th}$ to September $27^{th}$, using two kind of seismometers:

\begin{itemize}
\item a broadband triaxial Trillium 240 seismometer made by Nanometrics Inc \cite{Trillium-manual} that some of the authors  LN, MP and FR) used also to characterise other sites, which are candidates to host the Einstein Telescope  3G-GW detector; 
\item  SM3 seismometers developed by the Russian Academy of Sciences based on a pendulum with negative feedback\cite{CM-3}. One of them was initially installed close to the Trillium 240 in the PK14 station, then  moved to the other stations.
\end{itemize}
The Trillium 240 seismometer was installed on the floor of the OGRAN laboratory, which is cemented to the bedrock of the tunnel, horizontally oriented towards the north direction and leveled. After the installation the sensor was left in position for about 24 hours to reach the thermal equilibrium with the local environmental temperature.

\noindent The Trillium 240 was connected to a DAQ Taurus \cite{Taurus-manual} to record the acquired data. 
The frequency response of the three channels is nearly flat in the microseismic band, and it rolls off at 40db/decade at lower frequencies. The analog output of the sensor is filtered using a first order low pass anti-alias filter before being sampled at this frequency. This data is later low pass filtered and decimated, using a 3 to 4 stage FIR filter, to the output sample rate of 40 Hz.

\begin{figure}[h!]
\centering
\includegraphics[width=1\textwidth]{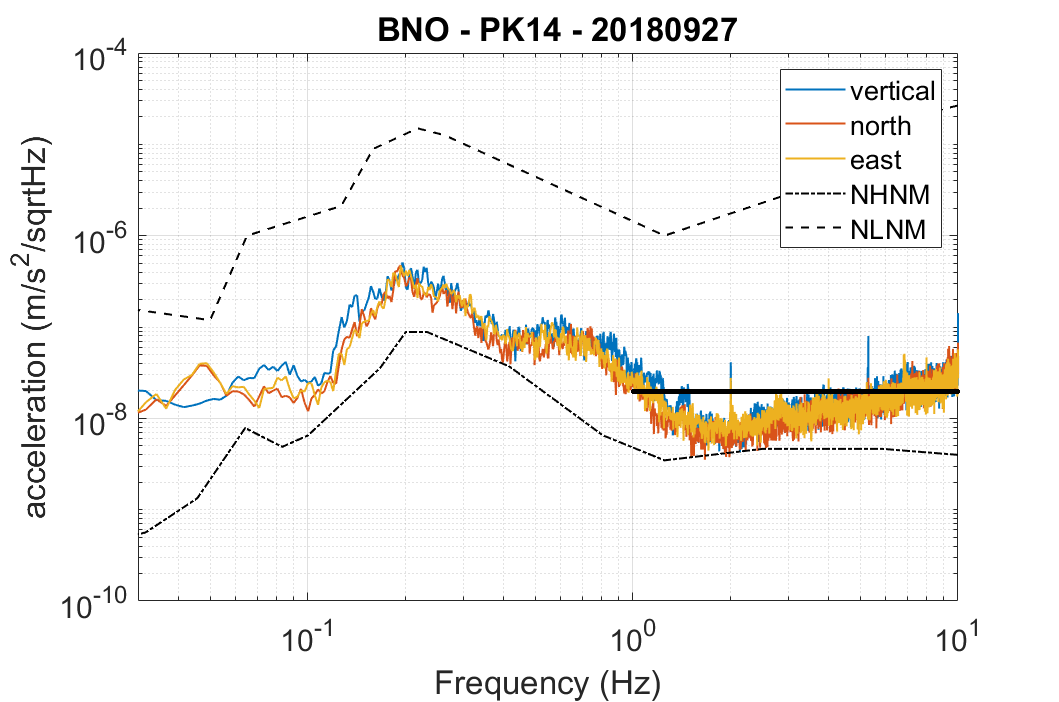}
\caption{Amplitude spectral density of the acceleration (median) along the vertical, E-W and N-S axis measured in OGRAN laboratory (near PK14) at BNO during the night of $27^{th}$ of September 2018. The solid black line is the ET requirement from \cite{ETdesign2011}}
\label{fig:ac_spectrum_27}
\end{figure}

In figure \ref{fig:ac_spectrum_27} we show the acceleration spectrum along three axis measured during the night between the $26^{th}$ and the $27^{th}$ of September 2018: the most prominent feature is the secondary oceanic microseismic peak at $f=0.195$ Hz \cite{longuet,Cessaro}, with a knee between $0.4$ Hz and $1$ Hz that is likely related to the waves in seas with a smaller characteristic wavelength than oceans \cite{Naticchioni}, \textit{e.g.}\ the Black and Caspian seas. Above $1$ Hz both the horizontals and vertical spectra are close to the minimum level given by the NLNM. From $\sim1$ to $\sim6$ Hz this spectrum is below the Einstein Telescope project seismic requirement\cite{ETdesign2011}. Comparing the spectrum with the previous measurements made in 2013, and taking in mind the square factor between amplitude and power spectral densities, it is possible to note an excess seismic noise above 2 Hz in the previous measurements. This could be due to a transient noise in the BNO infrastructure and/or to the measurement setup adopted at that time.

\begin{figure}[h!]
\centering
\includegraphics[width=1\textwidth]{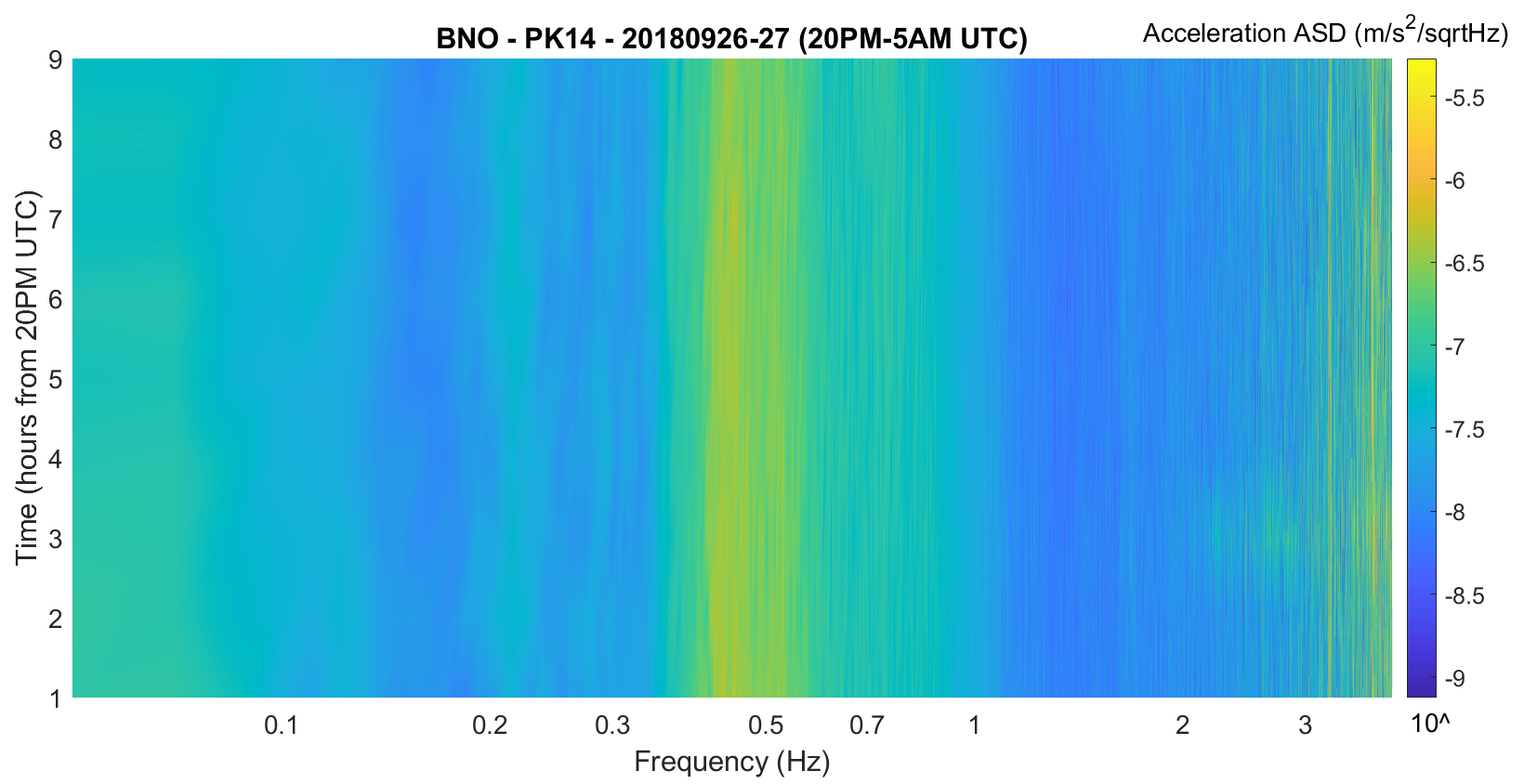}
\caption{Acceleration spectrogram along the vertical axis measured in station PK14 at BNO in the night between the $26^{th}$ and the $27^{th}$ of September 2018 with the Trillium 240 seismometer and a Taurus digitizer with 40Vpp input range}
\label{fig:ac_spectrogram_27}
\end{figure}

In figure \ref{fig:ac_spectrogram_27} we report the spectrogram obtained from one hour-long averaged spectra measured between $20$ PM of the $26^{th}$ and $5$ AM of the $27^{th}$ of September. The spectrogram shows the stability of the microseismic peaks and the absence of notable transients during the measurement.   

\section{Microseismic coherence across the tunnel}
Microseismic peaks are produced by the action of sea waves in the shallow waters of coastal regions and by ocean swell and standing waves in off-shore locations\cite{longuet,Cessaro, Cannata}.
During the period of June 2013 we had dedicated   data taking with the purpose to evidence of the existence of space coherence of the microseismic noise in the tunnel. This data taking was relatively short (few hours) and we focused the attention just on the vertical microseismic displacement. The noise spectrum of data sampled  at 100 Hz, have been  analysed up to maximum frequency of 10 Hz. We note that the noise was rather stationary during the short  period of data taking, and no glitches or special event have been detected. 

\noindent We assumed as base reference,   the PK38 station (3800 m from the main entrance of the tunnel) and  the synchronisation of the data streams collected in the other two  stations were corrected to the respect of the reference station.

\begin{figure}[h!]
\centering
\includegraphics[width=0.9\textwidth]{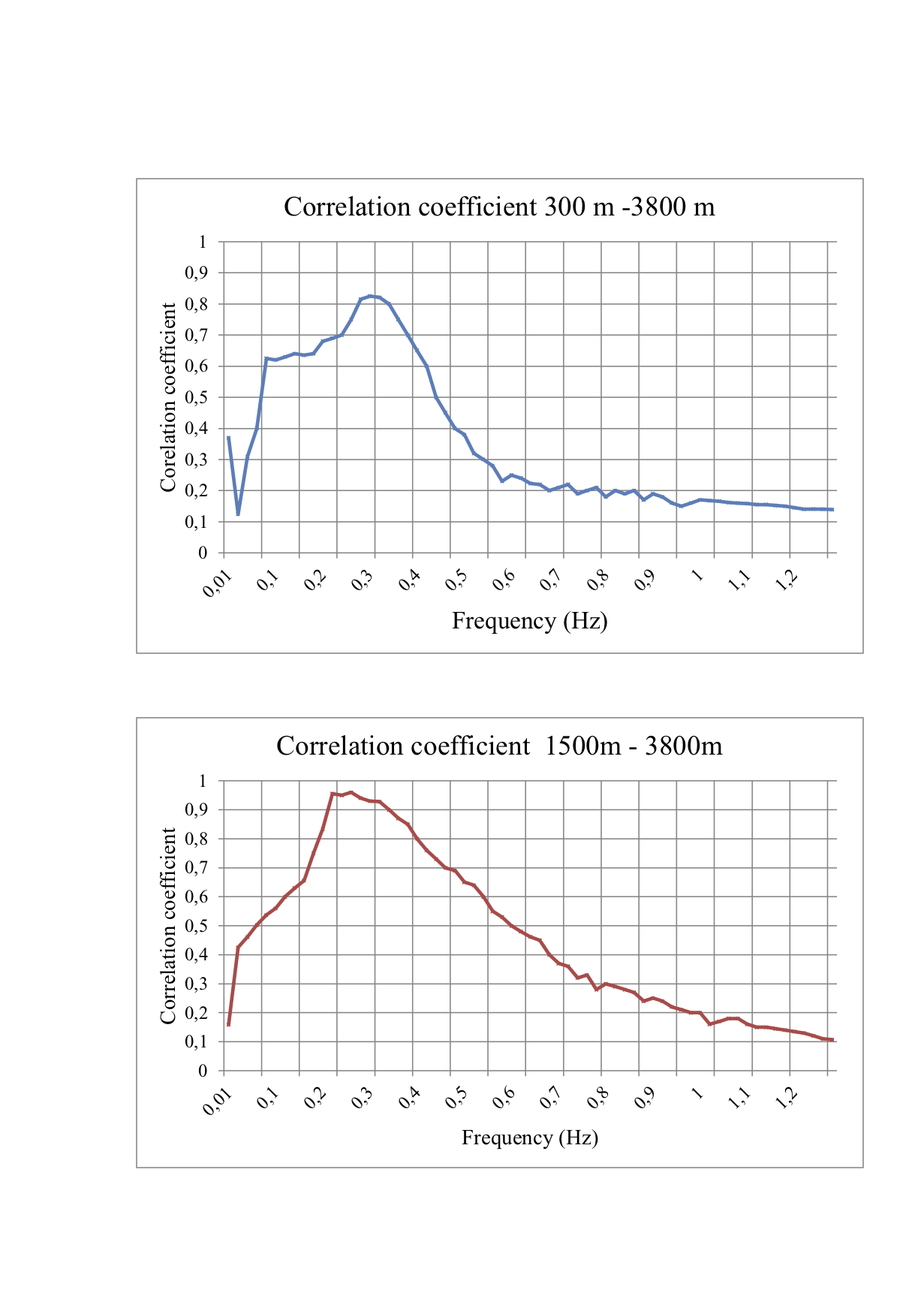}
\caption{Microseismic correlation in function of the frequency measured between the stations Pk3- Pk38 (top view) and  PK14-PK38 ( bottom view)}
\label{fig:correlations}
\end{figure}

\noindent Then, we computed the  the coherence in the  frequency range  from $10^{-3}$ to 10 Hz both in the case of PK3-PK38 and of PK14-PK38.
The two coherence plots in function of the  frequency are shown in figure \ref{fig:correlations}

\noindent It is evident from the two plots that the coherence is significant  in the frequency range $0.2~-~0.4$ Hz,  the region dominated by the microseismic peak. In other frequency intervals the correlation is weak or absent. 

\noindent Similar result has been obtained using  data collected  in 2018. We  monitored still the vertical microseismic displacement  using synchronised data of  two SM3 sensors located at PK3 and Pk38 data. 
\begin{figure}[h!]
\centering
\includegraphics[width=0.8\textwidth]{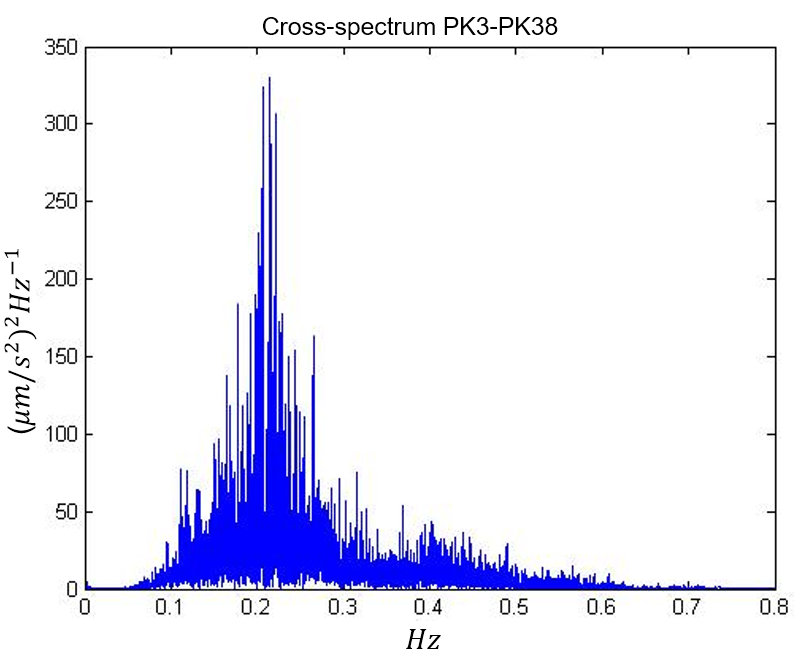}
\caption{Cross spectrum of the data collected in 2018  between the stations Pk3-Pk38}
\label{fig:correlations-3}
\end{figure}
The measurement confirms  the presence of a peak in the same frequency range for the cross correlation between the PK3 and PK38 stations. 

\section{Newtonian noise evaluation}

In a deep underground laboratory as BNO, the contribution of surface Rayleigh waves to the Newtonian noise (NN) is expected to be negligible, while that of compressional body waves is still important \cite{Amman, Harms1}. Under these assumptions, it is possible to calculate the NN projection produced by body waves, as discussed in \cite{Harms2} and \cite{DeRosa}. In a first approximation, we can assume the detector caverns as spherical cavities with a radius smaller than the characteristic seismic wavelengths, and the detector test masses as placed at center of these cavities \cite{Harms2}. Moreover, we can assume that the NN contribution to the interferometer test masses is not coeherent \cite{Hild2011}, and an equal distribution of the displacement power among the three body wave polarizations \cite{Amman}. 
With these assumptions, the NN projection of the seismic body waves background is expressed by:

\begin{equation}
    \tilde{h}_{NN}(f)=\frac{4\pi G \rho_0}{3}\frac{2\sqrt{2}}{L}\frac{1}{(2\pi f)^2}\tilde{x}(f)
\label{eq:NNapprox}    
\end{equation}

\noindent where $\rho_0$ is the average density of the local rocks ($\rho_0=2800$ kg/m$^3$), $L$ is the interferometer arms length ($L=10$ km in ET) and $\tilde{x}(f)$ is the seismic amplitude spectral density (ASD) measured at the site.

\begin{figure}[h!]
\centering
\includegraphics[width=1\textwidth]{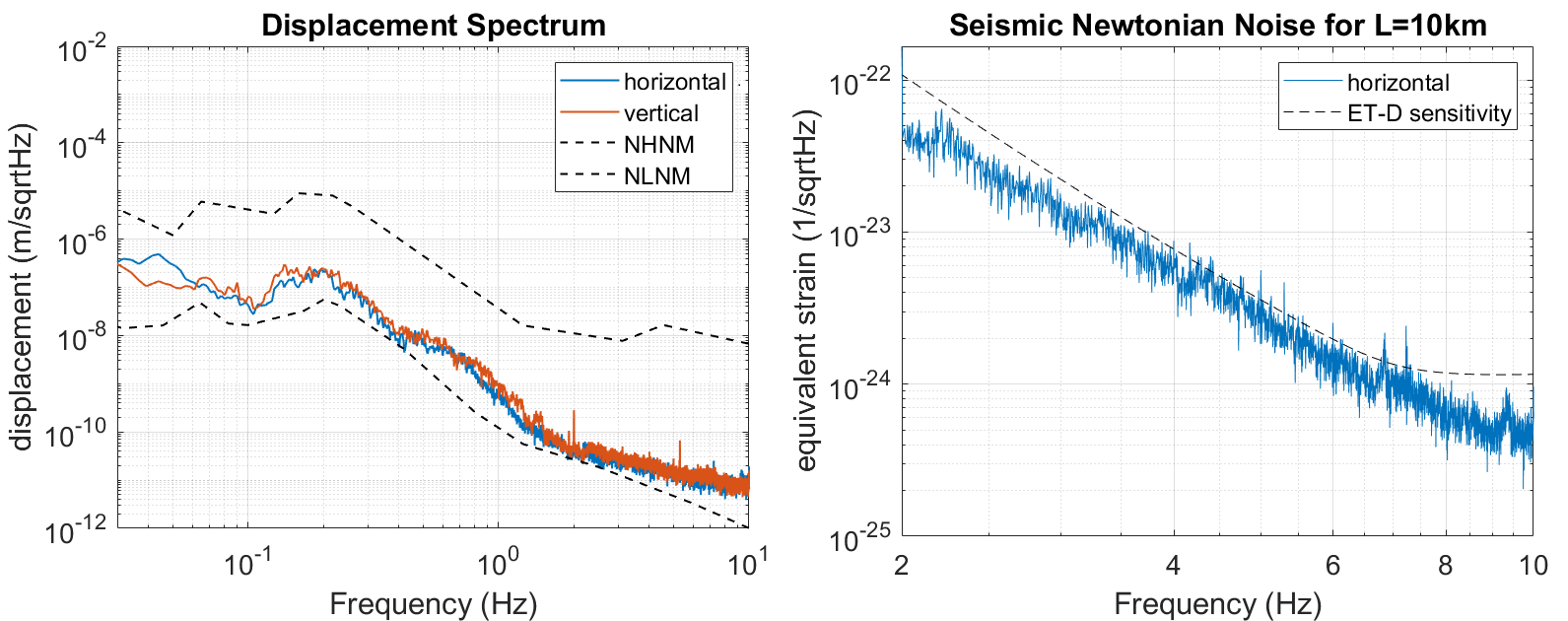}
\caption{Left: displacement amplitude spectral density (median) of the data acquired in the 2018 measurement campaign at PK14 (850 m underground). Right: Newtonian noise projection equivalent strain calculated from the horizontal displacement spectrum compared to the design sensitivity of ET, in the 2-10 Hz band}
\label{fig:NN-computed}
\end{figure}
\noindent In figure \ref{fig:NN-computed} we show the displacement ASD measured in 2018 at PK14 and the Newtonian Noise projection obtained from this data using equation \ref{eq:NNapprox}. It can be noticed that in the low frequency region for a 3G-GW (2-10Hz) the seismic noise background at BNO-PK14 (850m underground) is close to the Peterson's New Low Noise Model. The expected seismic NN, based on this data, is below the ET\cite{ETsensitivity} in almost all the $2~-~10$ Hz band.

\section{Discussion and conclusion}
 We have presented the experimental results of our campaign on seismic measurements carried on in the Baksan Neutrino Observatory (BNO). The two series of measurements has been taken at a distance of $\sim5$ years with different seismometers and show similar seismic PSDs. The spectra reported in figure \ref{fig:2013spectra} show clearly that the  deeper location (PK34 station) is more quite  above 1 Hz, a frequency range interesting for the site qualification of a 3G-GW detector.
Down to $1$ Hz the spectrum is affected by the sea motion: it is well known that closed seas and big lakes cause a characteristic microseismic peak at higher frequencies than that related to the oceans, due to the shorter typical wavelength in smaller bodies of water. In the case of BNO we can see an additional peak at $\sim 500-600$ mHz that can be interpreted as produced by Caspian Sea and/or the Black Sea. Indeed, the correlation of microseism and sea waves is a known effect also observed in other locations\cite{Cannata}, e.g.\ in the ET candidate site in Sardinia\cite{Naticchioni,MDG}, where a microseismic peak is well correlated with waves produced by the western sector of the Mediterranean sea\cite{MDG}. No evidence of  the dynamic effects due to magma movements of the near volcano have been detected at BNO.

 In addition, we have computed the potential contribution of the expected Newtonian Noise using the seismic data acquired in 2018 at the PK14 station (about 850m underground): this NN projection is below the Einstein Telescope design sensitivity in almost all the band from 2 to 10 Hz.
 
 Finally, we note  that the cross spectrum of  seismometers located in different positions along the tunnel  shows  the main peak at 0.21 Hz. The interpretation of this result is not straightforward: the correlation  on the microseismic spectrum in the range $0.1-0.4$ Hz  can be simple due to the secondary (dominant) microseismic peak produced by standing waves in the oceans, usually at $f\sim 0.2$ Hz, or it may also show a resonance effect of the sound propagation in the tunnel. Future measurements with infra-sound microphones could permit to asses the correct interpretation of this result.

\section*{Acknowledgments}
We thank the Direction of Baksan Neutrino Observatory for their kind hospitality during the data taking. Russian participants were supported by the national grant   RFBR 19-29-11010.  The support of Sapienza Grande Progetto di Ateneo 2017, Amaldi ResearchCenter (MIUR program "Dipartimento di Eccellenza" CUP:B81I18001170001) and of Istituto Nazionale di Fisica Nucleare is gratefully acknowledged.  
\\
\\
\noindent \textbf{Data Availability Statement.} This manuscript has associated data in a data repository. [Authors’ comment:
The 2018 measurement campaign data is stored in repository and is available upon request. Please contact the corresponding author.]


\end{document}